\documentstyle{article}

\def\Journal#1#2#3#4{{#1} {\bf #2}, #3 (#4)}

\def\PRD{{\em Phys. Rev.} D}
\def\ZPC{{\em Z. Phys.} C}

\def\be{\begin{equation}}
\def\ee{\end{equation}}
\def\bea{\begin{eqnarray}}
\def\eea{\end{eqnarray}}

\begin{document}
\pagestyle{plain}
\title{FORM FACTORS IN AN ALGEBRAIC MODEL OF THE NUCLEON}

\author{R. BIJKER\\
Instituto de Ciencias Nucleares, U.N.A.M., A.P. 70-543,\\
04510 M\'exico, D.F., M\'exico\\
\and
A. LEVIATAN\\
Racah Institute of Physics, The Hebrew University,\\
Jerusalem 91904, Israel}
\date{September 8, 1995}
\maketitle

\begin{abstract}
We study the electromagnetic form factors of the nucleon in a
collective model of baryons. In an algebraic approach to hadron
structure, we derive closed expressions for both elastic and transition
form factors, and consequently for the helicity amplitudes that
can be measured in electro- and photoproduction.
\end{abstract}

\section{Introduction}

Currently there is a lot of interest in the structure of the nucleon
and its resonances \cite{baryons}. The (re)measurement of the
electromagnetic form factors of baryon resonances forms an
important part of the experimental program at various facilities,
{\it e.g.} MAMI and CEBAF.

Extensive calculations of these observables were carried out
in the relativized quark potential model \cite{capstick},
in which a relatively small number of low-lying configurations
in the confining potential contribute significantly to the
baryon wave functions. On the other hand, flux-tube models,
soliton models, as well as some regularities in the observed
spectrum (linear Regge trajectories and parity doubling) hint
that an alternative, more collective type of dynamics may play a
role in the structure of baryons.

In this contribution we discuss the electromagnetic couplings
in a collective model using an algebraic approach for baryons.

\section{Algebraic model of the nucleon}

In \cite{bil} we introduced an algebraic model, in which
the nucleon has the string configuration of Figure~1. Its three
constituent parts are characterized by the internal degrees of freedom
of spin, flavor and color and by the two relative Jacobi coordinates,
$\vec{\rho}$ and $\vec{\lambda}$, and their conjugate momenta. For these
six spatial degrees of freedom we suggested to use a $U(7)$ spectrum
generating algebra whose building blocks are six dipole
bosons, $b^{\dagger}_{\rho,i}$ and $b^{\dagger}_{\lambda,i}$
($i=x,y,z$), and an auxiliary scalar boson, $s^{\dagger}$.
For a system of interacting bosons the model space is spanned by the
symmetric irreps $[N]$ of $U(7)$, which contains the oscillator shells
with $n=0,1,2,\ldots,N$. Here $N$ is the conserved total number of bosons.
The full algebraic structure is obtained by combining the geometric part,
$U(7)$, with the internal spin-flavor-color part,
$SU_{sf}(6) \otimes SU_c(3)$.

For the nucleon (isospin $I=1/2$) and delta ($I=3/2$) families of
resonances the three strings of Figure~1 have equal
length and equal relative angles. Hence this configuration is an
oblate top and has $D_{3h}$ point group symmetry.
The classification under $D_{3h}$ is equivalent to
the classification under permutations and parity.
States are characterized by
$(v_1,v_2);K,L_t^{P}$, where $(v_1,v_2)$ denote the vibrations
(stretching, bending); $K$ denotes the projection of the rotational
angular momentum $L$ on the body-fixed symmetry axis, $P$ the parity and
$t$ the symmetry type of the state under $D_3$ (a subgroup of $D_{3h}$
isomorphic to the $S_3$ permutation group).
The symmetry type of the geometric part must be the same as that
of the spin-flavor part (the color part is antisymmetric). Therefore,
one can use the representations of either $D_3$ or $SU_{sf}(6)$
to label the states: $A_1 \leftrightarrow 56$,
$A_2 \leftrightarrow 20$, $E \leftrightarrow 70$.

In \cite{bil} we used a $D_3$-invariant mass operator consisting of
spatial and spin-flavor contributions to obtain a description of the mass
spectrum of nonstrange baryons with a r.m.s. deviation of 39 MeV.
In this collective model of the nucleon baryon resonances are interpreted
as vibrations and rotations of an oblate symmetric top.
The corresponding wave functions, when expressed in a harmonic oscillator
basis, are spread over many shells and hence are truly collective.

\section{Electromagnetic form factors}

Helicity amplitudes and form factors can be measured in photo- and
electroproduction of baryon resonances. In the nonrelativistic limit
the transverse coupling of the photon field to the three constituent
parts is given by a magnetic and an electric contribution
\bea
{\cal H} &=& 6 \sqrt{\pi/k_0} \, \mu_3 e_3 \left[ k s_{3,+} \hat U
- \hat T_+/g_3 \right] ~.
\eea
Here $\vec{k}=k\hat z$ is the photon momentum, $k_0$ the photon
energy and $s_3$, $e_3$, $g_3$, $m_3$, $\mu_3=eg_3/2m_3$ denote
the spin, charge, $g$-factor, mass and the scale magnetic moment of the
third constituent, respectively.
The operators $\hat U$ and $\hat T_+$ only act on the spatial part
of the baryon wave function. In an algebraic treatment they are
given by
\bea
\hat U &=& \mbox{e}^{ -i k \beta \hat D_{\lambda,z}/X_D } ~,
\nonumber\\
\hat T_+ &=& \frac{i m_{3} k_0 \beta}{2 X_D} \left( \hat D_{\lambda,+} \,
\hat U + \, \hat U \, \hat D_{\lambda,+} \right) ~, \label{emop}
\eea
where $\hat D_{\lambda,m} = (b^{\dagger}_{\lambda} \times s -
s^{\dagger} \times \tilde b_{\lambda})^{(1)}_m$ is the dipole operator
in $U(7)$ which has the same transformation properties
as the Jacobi coordinate $\lambda_m$.
The coefficient $X_D$ is a normalization
factor and $\beta$ represents the scale of the coordinate.

Since $\hat D_{\lambda}$ is a generator of the algebra of $U(7)$,
the matrix elements of $\hat U$ are representation
matrix elements of $U(7)$, {\it i.e.} generalized Wigner
${\cal D}$-matrices. By making an appropriate
basis transformation they can be obtained exactly. However, in the
limit of $N \rightarrow \infty$ (infinitely large model space)
they can also be derived in closed form. This derivation consists of
several steps. The rotational states $|K,L,M \rangle$ of the ground
state band of the oblate symmetric top can be obtained by projection
from an intrinsic (or coherent) state
\bea
|K,L,M\rangle &=& \sqrt{\frac{2L+1}{8\pi^2}} \int d\Omega \,
{\cal D}_{MK}^{(L) \ast}(\Omega) \, {\cal R}(\Omega) \, |N,R\rangle ~,
\nonumber\\
|N,R\rangle &\propto& \left( s^{\dagger} + R (b^{\dagger}_{\lambda,x}
+b^{\dagger}_{\rho,y})/\sqrt{2} \right)^N \, |0\rangle ~.
\eea
The coefficient $R$ appears in the mass operator and is associated
with the size of the string.
Next we construct states with good $D_3$ symmetry by taking
the linear combinations
\bea
|\psi_1\rangle &=& \frac{1}{\sqrt{2(1+\delta_{K0})}} \,
\left[ (-)^L |K,L,M\rangle + |-K,L,M\rangle \right] ~,
\nonumber\\
|\psi_2\rangle &=& \frac{i}{\sqrt{2(1+\delta_{K0})}} \,
\left[ |K,L,M\rangle -(-)^L |-K,L,M\rangle \right] ~. \label{wf}
\eea
For $K(\mbox{mod }3)=0$ the wave function $|\psi_1\rangle$ is symmetric
($A_1 \leftrightarrow 56$) and $|\psi_2\rangle$ antisymmetric
($A_2 \leftrightarrow 20$), whereas for $K(\mbox{mod }3) \neq 0$
the wave functions $|\psi_1\rangle$ and $|\psi_2\rangle$ are the two
components of the mixed-symmetry doublet ($E \leftrightarrow 70$).
Eq.~(\ref{wf}) is consistent with the choice of geometry in $|N,R\rangle$.
Finally, the matrix elements of $\hat U$ and $\hat T_+$ can be derived
in closed form in the $|K,L,M\rangle$ basis. For example,
for $\hat U$ we find
\bea
\langle K,L,M | &\hat U& | K^{\prime},L^{\prime},M^{\prime} \rangle
\;=\; i^{K-K^{\prime}} \, \sqrt{(2L^{\prime}+1)/(2L+1)}
\nonumber\\
&& \sum_{\lambda} \frac{1}{2}
\left[ 1+(-1)^{\lambda+K-K^{\prime}} \right] \,
(2\lambda+1) \, j_{\lambda}(k\beta) \,
\langle L^{\prime},M^{\prime},\lambda,0|L,M \rangle
\nonumber\\
&&
\langle L^{\prime},K^{\prime},\lambda,K-K^{\prime}|L,K \rangle \,
\frac{\sqrt{(\lambda+K-K^{\prime})!(\lambda-K+K^{\prime})!}}
{(\lambda+K-K^{\prime})!!(\lambda-K+K^{\prime})!!} ~,
\eea
where we have used that in the large $N$ limit the intrinsic matrix
element becomes diagonal in the orientation $\Omega$ of the condensate.
For $\hat T_+$ we find a similar expression in terms of spherical
Bessel functions.

In the collective model discussed here, these spatial matrix elements
are folded with a distribution function
$g(\beta)=\beta^2 \exp(-\beta/a)/2a^3$
for the charge and the magnetization along the string.
With this distribution we reproduce the
observed dipole form for the electric form factor of the proton
$G^p_E=1/(1+k^2a^2)^2$~.
The scale parameter $a$ can be determined from the proton charge radius.
The helicity amplitudes for a given baryon
resonance can be obtained by combining the spatial contribution
with the appropriate spin-flavor matrix elements \cite{bil}.
For example, the proton helicity amplitudes for the
N(1520)$D_{13}$ resonance are given by
\bea
A_{1/2} &=& 2i \mu \, \sqrt{\pi/k_0} \,
\left[ m_3 k_0 a/g_3 - k^2 a \right]/(1+k^2a^2)^2 ~,
\nonumber\\
A_{3/2} &=& 2i \mu \, \sqrt{3\pi/k_0} \,
m_3 k_0 a/g_3(1+k^2a^2)^2 ~.
\eea
Just as in the harmonic oscillator quark model,
the asymmetry parameter
$A=(A_{1/2}^2-A_{3/2}^2)/(A_{1/2}^2+A_{3/2}^2)$ changes rapidly from
$-1$ to $+1$ with increasing momentum transfer. For small values of $k$
the helicity-$1/2$ amplitude is small because of the canceling
contributions of the electric and magnetic terms, whereas
for large values of $k$ the contribution of the magnetic term
dominates. However, in the collective model the helicity amplitudes
fall as powers of the momentum transfer, in contrast to the
exponential decrease of harmonic oscillator form factors.
It is important to note that this property holds for all baryon
resonances. Further improvements can be obtained by the
introduction of spin-flavor symmetry breaking and the stretching
of the strings \cite{bil2}.

\section{Summary and conclusions}

In conclusion, we have discussed electromagnetic form factors
of nonstrange baryon resonances in the context of a collective
model of the nucleon. Collective form factors are obtained by
folding with a probability distribution, which is determined
by the elastic form factor of the proton.
Within the assumptions of the model the transition form factors
to all other nonstrange baryon resonances are derived in closed
form and are predicted to drop as powers of the momentum transfer.
Such predictions of electromagnetic properties
can be tested in future photo- and electroproduction experiments.

\section*{Acknowledgements}

This work is supported in part by CONACyT, M\'exico under project
400340-5-3401E, DGAPA-UNAM under project IN105194 and the Basic Research
Foundation of the Israel Academy of Sciences and Humanities, and
by grant No. 94-00059 from the United States-Israel Binational
Science Foundation (BSF), Jerusalem, Israel.

\begin{figure}[h]
\caption[]{Collective model of baryons and its idealized string
configuration (the charge distribution of the proton is shown as
an example).}
\end{figure}


\begin{thebibliography}{99}

\bibitem{baryons}
See e.g. {\em Baryons `92}, Ed. M. Gai
(World Scientific, Singapore, 1993);
{\em Sixth Workshop on Perspectives in Nuclear Physics at Intermediate
Energies}, Eds. S. Boffi, C. Ciofi degli Atti and M. Giannini
(World Scientific, Singapore, 1994).

\bibitem{capstick}
M. Warns, H. Schr\"oder, W. Pfeil and H. Rollnik,
\Journal{\ZPC}{45}{627}{1990};
F.E. Close and Z. Li,
\Journal{\PRD}{42}{2194}{1990};
Z. Li and F.E. Close,
\Journal{\PRD}{42}{2207}{1990};
M. Warns, W. Pfeil and H. Rollnik,
\Journal{\PRD}{42}{2215}{1990};
S. Capstick,
\Journal{\PRD}{46}{2864}{1992};
S. Capstick and B.D. Keister,
\Journal{\PRD}{51}{3598}{1995}.

\bibitem{bil}
R. Bijker, F. Iachello and A. Leviatan,
\Journal{\em Ann. Phys. (N.Y.)}{236}{69}{1994}.

\bibitem{bil2}
R. Bijker, F. Iachello and A. Leviatan, in preparation.

\end{thebibliography}
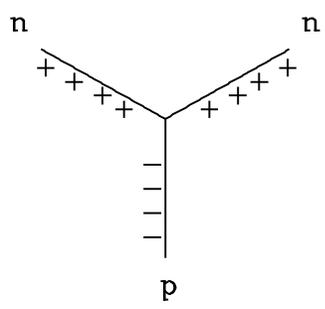
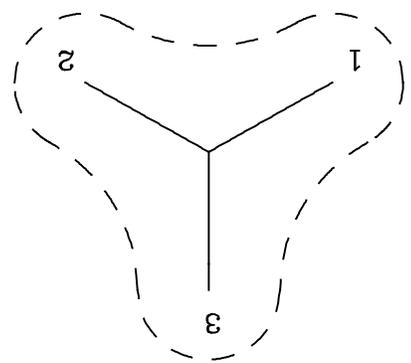
\end{document}